\documentclass[aps,prx,floatfix,superscriptaddress,noshowpacs,twocolumn]{revtex4-2}
\usepackage{graphics,graphicx,epsfig,amsmath,amssymb,wasysym,epstopdf}
\usepackage{hyperref}
\usepackage[dvipsnames]{xcolor}

\begin{document}
\title{Interplay between edge states and charge density wave order in the Falicov-Kimball model on a Haldane ribbon}

\author{Jan Skolimowski}
\email{jan.skolimowski@sissa.it}
\affiliation{International School for Advanced Studies (SISSA), Via Bonomea 265, I-34136 Trieste, Italy}

\begin{abstract}

To determine the impact of including edge states on the phase diagram of a spinless Falicov-Kimball model (FKM) on the Haldane lattice, a study of a corresponding ribbon geometry with zigzag edges is conducted. By varying the ribbon widths, the distinction between the effects connected to the mere presence of the edges and those originating from interference between the edge states is established. The local doping caused by the former is shown to give rise to a topologically trivial bulk insulator with metallic edge states. Additionally, it gives rise to a charge density wave (CDW) phase with mixed character of the subbands in various parts of the phase diagram. The local doping on the CDW instability is also addressed. Two additional gapless phases are found, caused by the edges but with stability regions depending on the width of the ribbon.


\end{abstract}

\maketitle

\section{Introduction}

The study of topological phases of matter has been a rapidly growing research direction in condensed matter physics over the past 30 years. The hallmark of non-trivial topology is the presence of topologically protected states at the boundary of a system. From zero dimensional Majorana states~\cite{oreg2010helical,mourik2012signatures} though chiral spin Hall edge states~\cite{PhysRevLett.95.226801,PhysRevLett.95.146802} to surface states in a crystalline topological insulator~\cite{PhysRevLett.106.106802,Hsieh2012} all of them carry quantized information about the bulk. Topological protection defends them against local perturbations such as lattice imperfections, making them a viable candidate for application in spintronics or quantum computing~\cite{He2019,doi:10.1073/pnas.1810003115}.

The stability of a non-trivial topological state becomes less obvious when strong Coulomb repulsion is involved. The effect of electron-electron interactions can vary from simple renormalization of the non-interacting bands to complete reconstruction of the band structure due to formation of a long-range ordered phase. 
The latter effect can be blocked by the lattice symmetries, promoting a topologically non-trivial ground state. However in some cases the order can circumvent these restrictions favoring toplogically trivial ground state, as in the case of the antiferromagnet in Kane-Mele-Hubbard model~\cite{PhysRevLett.106.100403,PhysRevB.85.205102}. 
This interplay of topology and various long-range orders has been explored in recent years in search of novel phases of matter. But interactions are not only detrimental to the non-trivial topology, they can also promote it. Multiple scenarios were put forward where this can take place, see Ref.~\cite{Rachel_2018} for a review. Interactions can for example suppress the existing order which excluded the non-trivial topology of the ground state~\cite{PhysRevB.87.235104}. Alternatively, longer-range electron-electron repulsion can induce a time-reversal breaking term and make the system topological~\cite{PhysRevLett.103.046811,PhysRevLett.100.156401}.
The result of the interplay between the long-range order and topology can become even more complicated when boundaries of the system are concerned. From the point of view topology, the edge state cannot exist without the bulk. Yet, if the bulk is too small the interference between the edge states can lead to their mutual cancellation~\cite{PhysRevB.90.035116}. From the point of view of formation of a long-range order the edges break the translation symmetry. In certain geometries the initial order can be made unstable if it becomes incommensurate with the new geometry. 
Despite a vast theoretical effort in exploring different scenarios where topology and long-range order compete in a finite geometry, the topic still presents open questions.

The goal of this work is to explore this interplay by studying the properties of the Falicov-Kimball model on a Haldane ribbon across the $U-T$ parameter space. 
At half-filling, when the total density of localized and itinterant electrons per site is one, the FKM on a bipartate lattice at $T=0$ is known to be unstable towards formation of a long-range order, the charge density wave (CDW) state~\cite{KENNEDY1986320,RevModPhys.75.1333}. On a Haldane lattice, which is not biparite, the CDW istability remains as long as the next-nearest neighbor hopping ($t^\prime$) is weak~\cite{Wojtkiewicz2006}.
The CDW phase of FKM has some unexpected features as a function of interaction strength and temperature that distinguishes it from the much better explored orders of Hubbard model. For example the mechanism through which it decays with $T$, the fact it is not affected by metallization, or that the metallic region has non-monotonic behavior as a function of $U$ and $T$, just to name a few. Despite few papers discussing some version of this model~\cite{PhysRevB.88.165132, PhysRevB.92.125102,PhysRevLett.122.126601}, none has investigated the scenario where both edge states and CDW instability are present. Thus, the aim of this paper is to explore the possible physical situations that can occur when both factors are at play.

\section{Main results and outline}

\begin{figure}[ht]
\begin{center}
\includegraphics[width=0.49\textwidth]{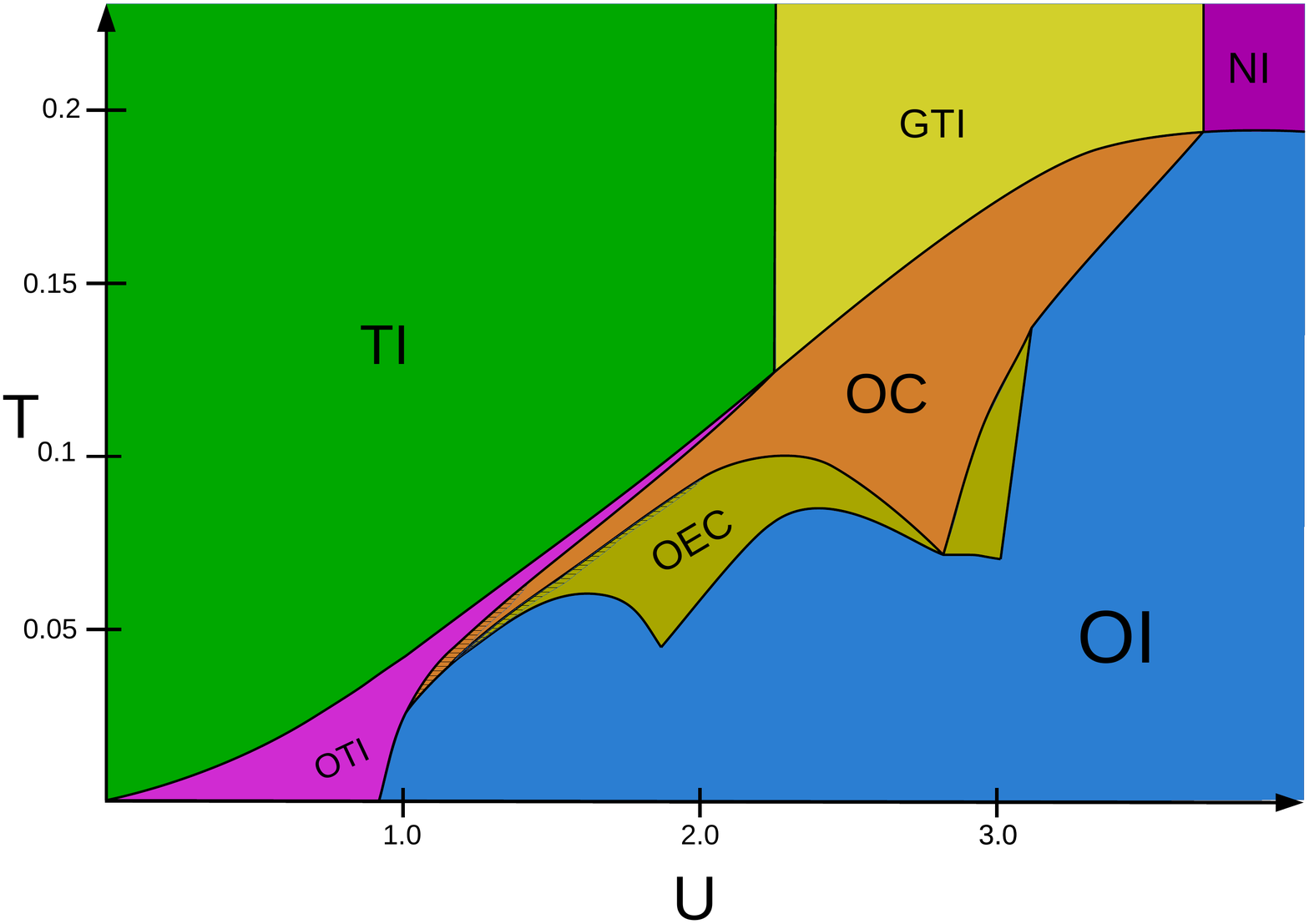}
\end{center}
\caption{A schematic depiction of the phase diagram of FKM on the Haldane ribbon of width L=16. It shows eight distinct phases: Topological insulator (TI), ordered TI (OTI), gapless TI (GTI), ordered insulator (OI), normal insulator (NI), ordered conductor (OC) and ordered edge conductor(OEC). The small dashed parts of the phase diagram at OTI-OC and OC-OEC transitions represent regions dominated by the effects connected to the finite width of the ribbon.}
\label{phase_diag}
\end{figure}

The main result of this paper is presented in Fig.~\ref{phase_diag}, which is a schematic depiction of the phase diagram for a ribbon of width $L=16$ and inifnite length. The size of the ribbon was chosen to be small enough to allow for interference between the edge states within the bulk, but not so small that the size effects dominate the physics. By examining the distribution of localized particles and the values of the mobile particles local spectra at the Fermi level one can distinguish between seven phases of this model: topological insulator (TI), ordered TI (OTI), normal insulator (NI), ordered conductor (OC), ordered insulator (OI), gapless TI (GTI) and ordered edge conductor (OEC). Each of the phases and the transition between them is discussed in the following sections. Further distintion inside the OC phase will be made through analysis of the subband symmetries, following Ref.~\cite{ActaPhysPolA.130.2}. The most striking consequence of introducing zigzag edges on the phase diagram comes in form of the OEC phase. It is characterized by a trivial bulk insulator with metallic edges, which originates from combination of low energy scattering subbands and local doping at the zigzag edges. This interplay is also responsible for a mixed, in subband asymmetry, nature of gaped phase inside the CDW region. The finite width of the ribbon allows to investigate the influence of penetration depth of the edge states on the stability of the non-trivial topological phase. A feature that proves to play a role at finite temperatures not only in extremely small systems. It is responsible for formation of a residual-metallic state (OC) at moderate $U$ upon the transitions from OTI to gaped CDW and from OC to OEC, cf. small dashed regions of the phase diagram~\ref{phase_diag} . The region of $U-T$ space this phase occupies, decays with the ribbon width. An analogous phase on the gapped side of this transition was not found due to different properties of the edge states in the topologically trivial state.
 
The paper is organized as follows. In the next section the model will be introduced and the method used to solve it will be discussed. Next, the analysis will begin with a section about $T=0$ evolution. It will cover not only the transition between the OTI and OI but also the consequences of finite width of the system. The following section will cover the same transition from OTI to TI but at finite temperatures. A transition that resembles that of an infinite lattice. This will be followed by the analysis of the change of the properties of the system as it evolves from OI to TI along a $U=1.2$ line, which is above the $T=0$ critical $U$ for the topological phase transition. The remaining sections will look at evolution of the system along different isothermal lines. This allows to explore the richness of phases originating from the interplay between different OI phases in FKM, one of them being the OEC. Lastly, a high-$T$ evolution will be discussed. This parameter regime is characterized by lack of order causing  the lines of transition points to be vertical.

\section{Model and method}
\begin{figure}[ht]
\begin{center}
\includegraphics[width=0.5\textwidth]{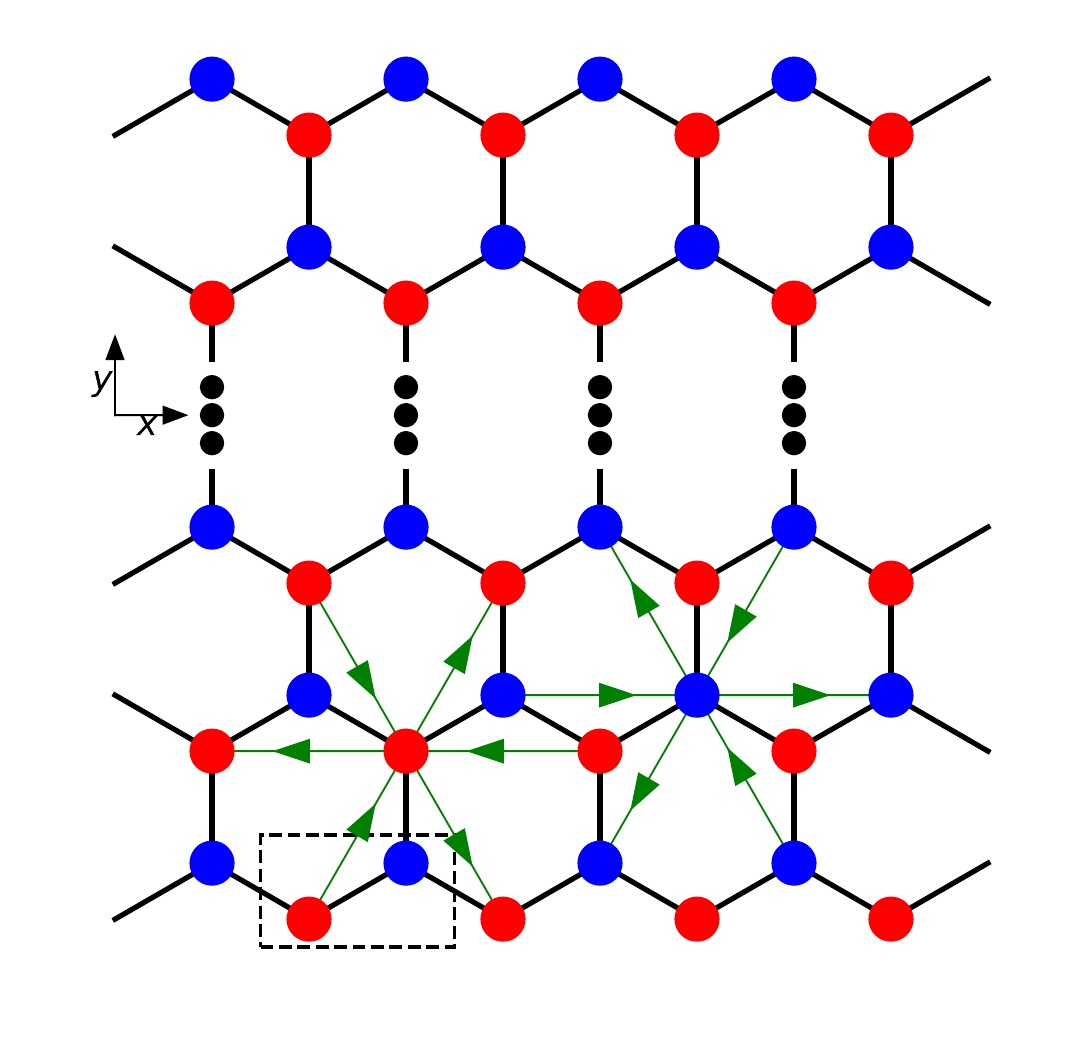}
\end{center}
\caption{Lattice scheme with blue and red circles indicating non-equivalent lattice sites on the honeycomb lattice. Black indicates the nearest neighbor hopping with amplitude $t=1$. Green line indicates NNN hopping with complex phase $\phi$, whose change is indicated with arrows. System is infinite in Cartesian $x$ direction and finite in $y$ direction }
\label{Fig0}
\end{figure}

The Hamiltonian of the FKM on a Haldane lattice in the second quantization form is given by:
\begin{multline}\label{H}
\mathcal{H}=-t\sum_{<i,j>} c^\dagger_{i}c_{j}+i t^\prime \sum_{<<i,j>>}\nu_{i,j} c^\dagger_{i}c_{j} + H.c.\\
+\sum_{i} U d^\dagger_{i}d_{i}c^\dagger_{i}c_{i}-\sum _i (E_d d^\dagger_{i}d_{}+\mu c^\dagger_{i}c_{i}),
\end{multline}
where $c^\dagger_i, d^\dagger_i$ are the creation operators of itinerant and localized spinless fermions. The first line in Eq.~\ref{H} describes a honeycomb lattice, with hopping amplitude $t$ between nearest neighbors, and with a complex next-nearest neighbor hopping $t^\prime $. Parameter $\nu_{i,j}=\pm 1$ distinguishes between a clockwise and anticlockwise movement of particles. Figure~\ref{Fig0} illustrates the Haldane lattice with its two atom unit cell (A and B type indicated as blue and red dots), directional hoppings, depicted as green lines with arrows, and non-directional hoppings indicated by black lines. The second line in Hamiltonian~(\ref{H}) describes the interaction between localized ($d$) and itinerant($c$) fermions of strength $U$ and $E_d, \mu$ set the chemical potential for $d$-electrons and $c$-electrons respectively.
In this work a semi-infinite version of Haldane lattice will be considered with a finite number of unit cells $L$ in $y$-direction, as shown in Fig.~\ref{Fig0}. This so-called ribbon geometry can host two main types of edges depending along which lattice vector it is cut~\cite{PhysRevB.54.17954}. Here the focus will be only on the zigzag type. In this geometry each edge host an excess of one type of atoms (A or B) leading to additional states even in the topological trivial case. This is not the case in the armchair edge, which sustains the sublattice symmetry.

The Hamiltonian~(\ref{H}) will be solved using the inhomogeneous dynamical mean-field theory (iDMFT). This method builds upon the DMFT assumption of locality of self-energy~\cite{DMFT-review} and applies it to systems with broken translation invariance in one or more directions. First introduced by Potthoff and Nolting~\cite{PhysRevB.59.2549}, this method has helped to understand the competition between lattice geometry and interactions. It was successfully applied to study both Hubbard model~\cite{PhysRevB.98.045133,PhysRevLett.122.036802,PhysRevB.70.241104,PhysRevB.93.165112} as well as the FKM~\cite{PhysRevB.70.195342,PhysRevB.73.205110,PhysRevB.85.205444}. The standard DMFT is a well established method, descibed in the context of the FKM in Refs~\cite{RevModPhys.75.1333,DMFT-review}. In the following paragraph the iDMFT procedure will be described with emphasis put on the steps characteristic to it. 

The crucial difference between the iDMFT and DMFT is that the local self-energy is not assumed to be the same at each lattice site. In the real space representation it is given by a (diagonal) matrix-valued function 
\begin{equation}
[\mathbf{\Sigma}(z)]_{\alpha,\beta}=\Sigma(z,\alpha)\delta_{\alpha,\beta},
\end{equation}
where $z=\omega+i0^+$ is the complex frequency, and $\alpha,\beta$ are lattice site indexes. Because iDMFT deals with inhomogeneous systems the local part of lattice Greens function cannot be calculated through Hilbert transform. Instead, it is obtained by directly evaluation of the resolvent. 
\begin{equation}
\mathbf{G}(z)=[\mathbf{G}^{-1}_0(z)-\mathbf{\Sigma}(z)]^{-1}
\end{equation}

To accelerate this process it is useful to take advantage of the remaining symmetries of the lattice. In the case of the ribbon geometry it is the translation invariance along one direction ($x$). Hence to calculate the local part of the lattice Greens function it is useful to work in mixed momentum $k$ ($x$-direction) and position ($y$-direction) representation.  
\begin{equation}
[\mathbf{G}(z)]_{\alpha,\alpha}=\frac{1}{2\pi}\int dk \left[\mathbf{G}(z,k)\right]_{\alpha,\alpha},
\end{equation}
where $k$ is the rescaled momentum by $\frac{\sqrt{3}}{2}$ and
\begin{equation}
\mathbf{G}(z,k)=[z\mathbf{I}-\mathbf{\Sigma}(z)-\mathbf{H}(k)]^{-1}.
\end{equation}

In the mixed representation the non-interacting part of the Hamiltonian~\ref{H} has a 2x2 block tridiagonal form due to two atoms per unit cell of the honeycomb lattice. The off-diagonal blocks of the Hamiltonian in this basis are given by 
\begin{equation}
[\mathbf{H}(k)]_{s,s+1}=\left[
\begin{array}{cc}
2t^\prime\sin(k) & 0 \\
t & -2t^\prime\sin(k)
\end{array}
\right]
\end{equation}
and the diagonal blocks
\begin{equation}
[\mathbf{H}(k)]_{s,s}=\left[
\begin{array}{cc}
-2t^\prime\sin(2k) -\mu & 2t\cos(k)\\
2t\cos(k) & 2t^\prime\sin(2k)-\mu
\end{array}
\right],
\end{equation}
where index $s$ enumerates the unit cells across the ribbon. The (block) tridiagonal form of the Hamiltonian in the mixed representation makes the calculation of the resolvent time efficient~\cite{Reuter_2012}. From the local parts of the lattice Greens function the iDMFT procedure follows the same steps as in the standard DMFT self-consistency loop. The non-interacting Greens function is extracted from the lattice Greens function through
\begin{equation}
\mathcal{G}^{-1}_0(z,\alpha)=[\mathbf{G}(z)]^{-1}_{\alpha,\alpha}+\Sigma(z,\alpha).
\end{equation}
The local density of localized particles is updated using this non-interacting Greens function through
\begin{equation}
\rho_d(\alpha)=n_{FD}(\tilde{E}_d(\alpha,T),T),
\end{equation}
where $n_{FD}$ is the Fermi-Dirac distribution function,$T$ is the temperature and
\begin{equation}
\tilde{E}_d(\alpha,T)=E_d-\int\frac{d\omega}{2\pi} n_{FD}(\omega,T)\mathrm{Im} \ln\left(\frac{1}{1-U\mathcal{G}_0(z,\alpha)}\right)
\end{equation}
is the effective local potential of the localized particles. Next, the local Greens function of the mobile particles is obtained using 
\begin{equation}\label{LDOS}
\mathrm{G}(z,\alpha)=\frac{1-\rho_d(\alpha)}{\mathcal{G}^{-1}_0(z,\alpha)}+\frac{\rho_d(\alpha)}{\mathcal{G}^{-1}_0(z,\alpha)-U}.
\end{equation}
Finally, the set of local self-energies is updated through
\begin{equation}
\Sigma(z,\alpha)=\mathcal{G}^{-1}_0(z,\alpha) -\mathrm{G}^{-1}(z,\alpha),
\end{equation} 
which is plugged-in to recalculate the lattice Greens function, closing the iDMFT loop. In addition at each steps a new $E_d$ is found such that the total density of localized particles fixed to $\rho_d=\frac{1}{N}\sum_\alpha \rho_d(\alpha)$, where $N$ is the total number of lattice sites.

Due to the translation invariance along $x$ direction, in the following the index $y$ will be used to enumerate the non-equivalent unit cells across the ribbon and additional index $\sigma=A,B$ will indicate the sublattice. The system size $L$ is defined as the number of $y$ values enumerating non-equivalent unit cells across the ribbon.  
In the rest of the paper $A_{\sigma,y}(\omega)$ will symbolize the local density of states(LDOS) or spectral function, which is minus imaginary part of the corresponding retarded local Greens function divided by $\pi$, at sublattice $\sigma$ at unit cell $y$ from the edge. The dispersion along the ribbon will be obtained through the analogous quantity obtained from the Greens function in the mixed, momentum and position, representation. The $\rho_d(\sigma,y),\rho_c(\sigma,y)$ will represent the density of localized electrons at sublattice $\sigma$ and $y$-th unit cell from the edge. To distinguish between the ordered and disordered phases the $d(y)= \rho_d(A,y)-\rho_d(B,y)$ will be used. The LDOS at Fermi level $A_{\sigma,y}(\omega=0)$ will play the role of order parameter for metallization. 

In this work $t=1$, the strength of the non-directional hopping, will set the energy scale. The next-nearest neighbor hopping will be fixed at $t^\prime=0.1t$. The localized particles density per lattice site will be set to $\rho_d=0.5$ and a fixed chemical potential $\mu=\frac{U}{2}$ will be assumed to keep $\rho_c=0.5$. Most results analyzed below are for a ribbon of width $L=16$. A finite-size scaling analysis revealed this to be the sweet-spot width to obtain results quantitatively close to the infinite lattice, but also to include the interference between the edge states.

\section{T=0}

At $T=0$ the density of localized electron is fixed to integer values. This makes the effects of interactions static and results in a sublattice splitting mass term of strenght $U$ felt by mobile particles~\cite{RevModPhys.75.1333}. It counters the gap inversion, cused by the directional $t^\prime$, and a topological phase transition (TPT) takes place for $U>U_{TPT}=6\sqrt{3}t^\prime$~\cite{PhysRevLett.61.2015}. This effect was fully captured within the Hartree approximation in previous studies~\cite{PhysRevB.92.125102}. For finite systems the critical interaction strength depends on the system size~\cite{PhysRevB.90.035116}. The tunneling of the edge states inside the ribbon reduces the gap width in the middle of the ribbon, which plays the role of the bulk. As a consequence the $U_{TPT}$ is reduced. This is why the quantum phase transistion between the OTI and OI in Fig.~\ref{phase_diag} takes place at $U<U_{TPT}\approx 1.0$. Increasing the ribbon width to $L=20$ already recovers the critical $U$ from the infinite lattice limit. This is in agreement with findings for the noninteracting Kane-Mele model~\cite{PhysRevB.90.035116}, which reduces to Haldane model if one spin channel is considered.

After the TPT the edges turn insulating but their zigzag geometry still manifest itself in form of additional bands. Instead of connecting the $K$ and $K^\prime$ points across the gap they now connect the two $k$-points within the same (valence/conduction) band. Combined with the CDW order the resulting band structure hosts an additional hole-band localized at one end of the ribbon and a electron-band at the other end. This is in cotrast to Kane-Mele model, where these bands have a time-reversal counterpart, which cancels the charge imbalance between the edges~\cite{PhysRevLett.95.146802}. 

\begin{figure}[ht]
\begin{center}
\includegraphics[width=0.5\textwidth]{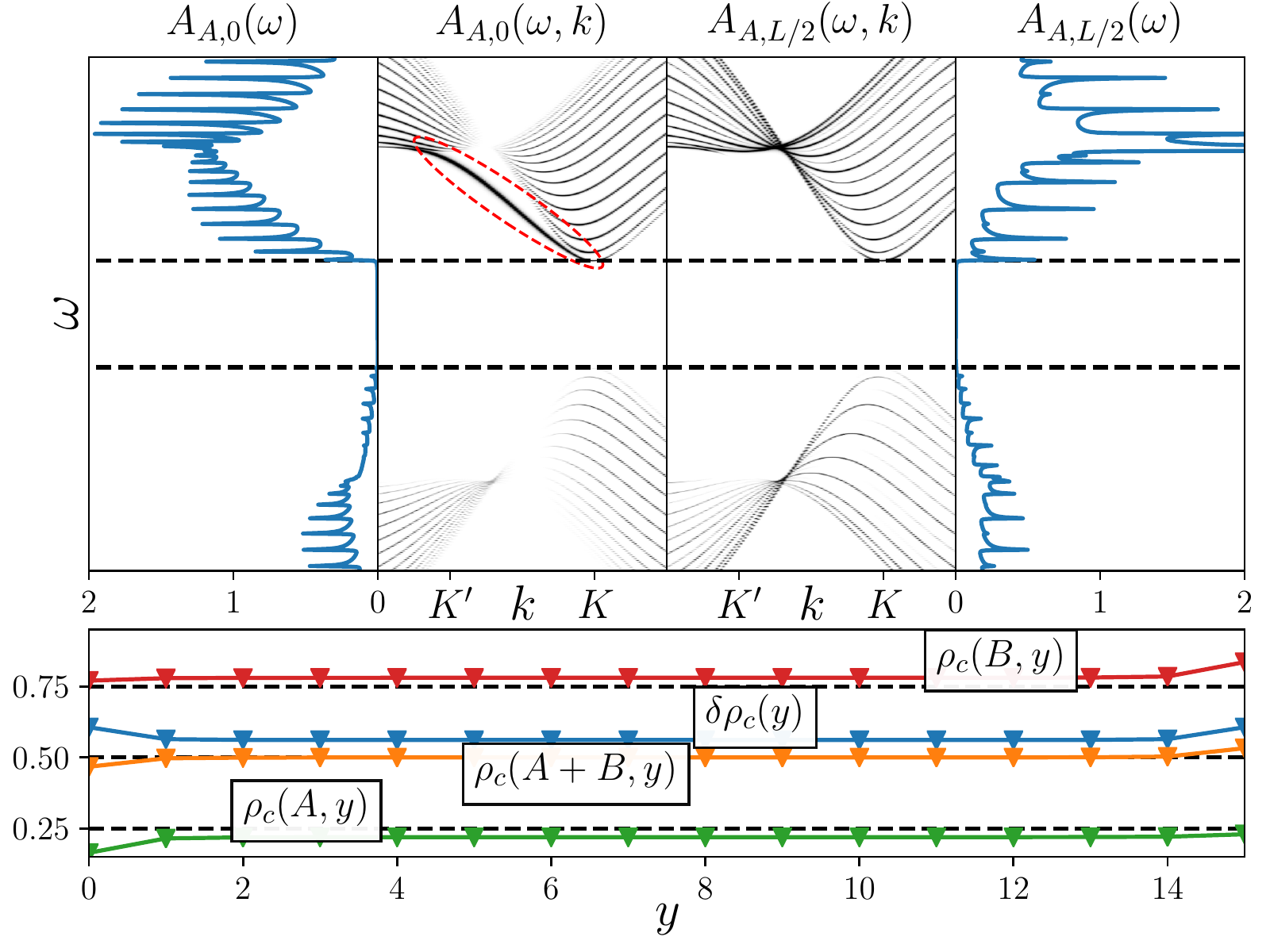}
\end{center}
\caption{Top row: dispersion and corresponding LDOS along the edge (left half) and along the middle of the ribbon (right half) for $U=2,T=0$. Dashed lines indicate the gap edges, and the red dashed ellipse in second from the left plot highlights the edges state after TPT. Bottom row: local density of mobile particles ($\rho_c(A,y),\rho_c(B,y)$), total density ($\rho_c(A+B,y)=0.5(\rho_c(A,y)+\rho_c(B,y))$) and polarization($\delta\rho_c(y)=\rho_c(A,y)-\rho_c(B,y)$) inside $y$ unit cell across the ribbon. }
\label{Fig3}
\end{figure}

As an illustration a dispersion along a single edge for $U=2$ and $L=16$ is shown in the second column in top plot of Fig.~\ref{Fig3}. The edge state after the TPT is highlighted by the red dashed ellipse. The dispersion at the other edge is not shown, due to symmetry it is just p-h transformed. The first column show the corresponding LDOS at the edge. As a reference point the third (fourth) column shows dispersion (LDOS) along (at) the middle of the ribbon. From comparison between the LDOSs one can see that the presence of the additional state at the edge triggers a transfer of spectral weight from below the gap to the additional states located above the upper gap edge. Despite that, the gap width is constatnt across the whole ribbon (black dashed lines). At $T=0$ the static correlations of FKM and integer valued density of localized particles does not allow for unit-cell-dependence of the gap width. This is in sharp contrast to the Hubbard-type interactions, where the correlations are dynamic and the reduced kinetic energy at the edges locally enhances the role of interactions~\cite{PhysRevB.59.2549,PhysRevB.98.045133,PhysRevLett.102.066806}. The fixed gap does not make the underlying lattice geometry insignificant, even in the topologically trivial phase.
On either side of the TPT the edge states tunnel into the ribbon in a particular way. They hybridize only with sites belonging to the same sublattice. This enhances the charge imbalance ($\delta\rho_c(y)=\rho_c(A,y)-\rho_c(B,y)$) within unit cells, which is most pronounced at the edges, cf. bottom panel of Fig.~\ref{Fig3}. Because one edge host an additional electron band and the other hosts a hole band the overall effect is transfer of the mobile electrons across the ribbon from one edge to the other. Such that the overall system remains half-filled. The effect is strongest in the vicinity of the TPT, when the edge states are well separated from the bulk bands. As $U$ grows the overlap in energy between bulk and edge bands becomes larger and the difference in mobile electron density between the edge unit cells decays.




\section{Non-zero temperature}

At finite temperatures the perfect checkerboard order becomes unstable and the oredr parameter $d(y)$ is no longer fixed to integer values. This is interpreted as the onset of temperature induced (annealed) disorder in the distribution of localized particles~\cite{PhysRevB.76.205109,PhysRevLett.117.146601}. The scattering off the disorder leads to formation of additional subbands inside the $T=0$ gap, located symmetrically around the Fermi level in the itinerant electron total (both sublattices) spectra. It is a general feature of the model, reported for various lattices~\cite{PhysRevB.76.205109,PhysRevB.77.035102,PhysRevB.89.075104}. It originates solely from the form of the local Greens function in the FKM, cf. Eq.~\ref{LDOS}. Depending on the which term has a larger numerator, the other term can be recast into a sum of scattering series from a potential of value $\pm U$ and the unperturbed environment described by the {\it majority} Greens function
\begin{multline}
\mathcal{G}(z,\alpha)=  
\left\{
\begin{array}{ll}
\left(1-\rho_d(\alpha)\right)\mathcal{X}+\rho_d(\alpha)\frac{\mathcal{X}}{1-U\:\mathcal{X}} \\
\rho_d(\alpha)\mathcal{X} +\left(1-\rho_d(\alpha)\right)\frac{\mathcal{X}}{1+U\:\mathcal{X}}\\
\end{array}
\right. 
\end{multline}
The top (bottom) recasting corresponds to the situation, where $\rho_d(\alpha)\approx 0$ ($\rho_d(\alpha)\approx 1$) and the unperturbed Greens function $\mathcal{X}$ is the solution from the perfectly ordered system $\mathcal{X}=\mathcal{G}_0(z,\alpha)$ ($\mathcal{X}=\left[\mathcal{G}^{-1}_0(z,\alpha)-U\right]^{-1}$). In both cases the position of the subbands correspond to the poles of the scattering series.
As the temperature grows the subbands inflate and eventually start to overlap, closing the gap. Multiple studies reported that this gapless CDW phase wedges between the ordered and disordered phases. 

\begin{figure}[ht]
\begin{center}
\includegraphics[width=0.5\textwidth]{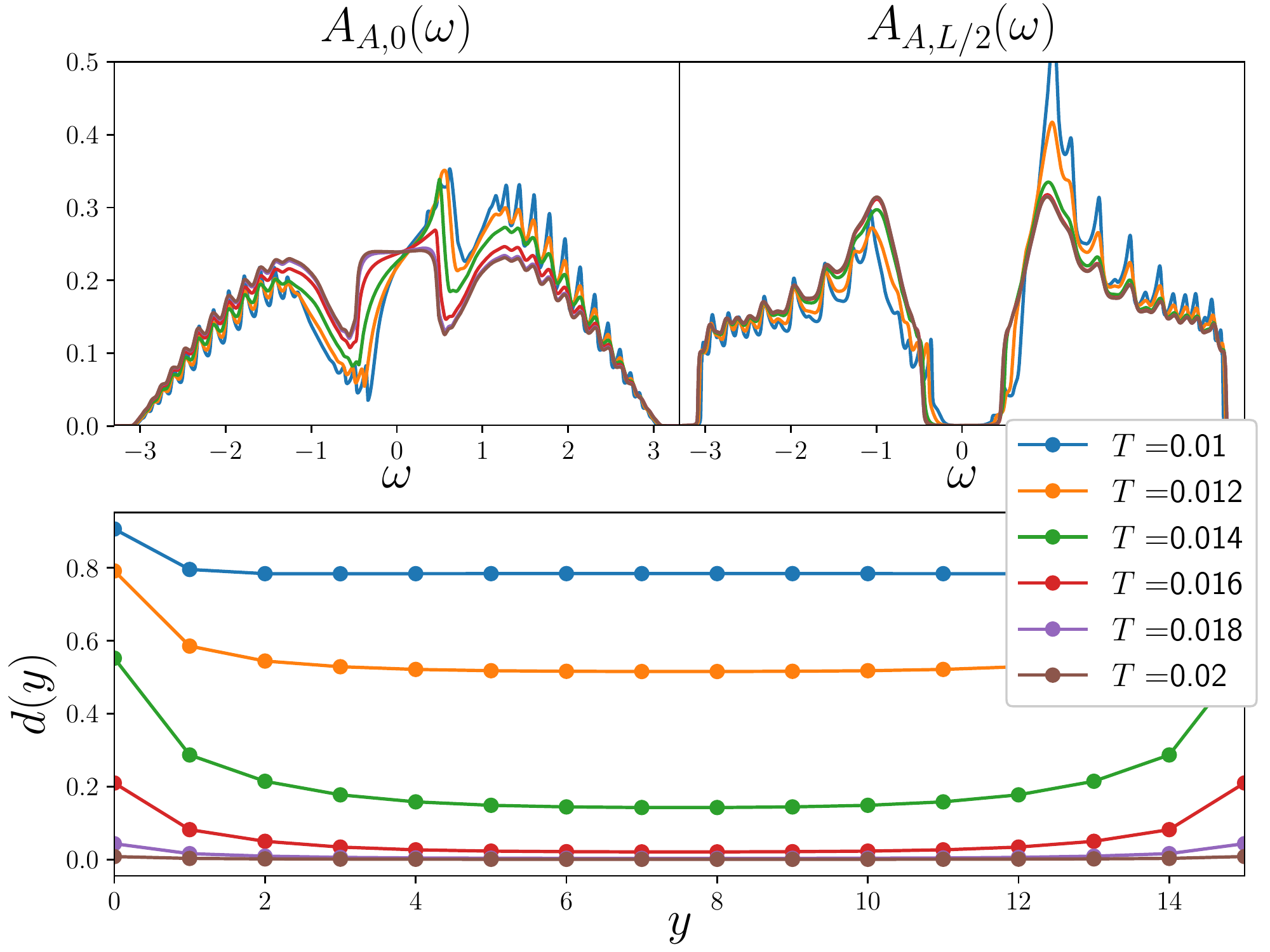}
\end{center}
\caption{Top row: Edge (left panel) and bulk (right panel) site-resolved LDOS for various $T$ across the ordered-disordered transition at $U=0.6$. Bottom row: the corresponding order parameter across the unit cells of the ribbon.}
\label{Fig4}
\end{figure}

On the Haldane lattice at $U<U_{TPT}$ this is not the case, which the results for ribbon geometry confirm. As long as the topology of the bulk does not change, the CDW order can decay without the intermediate metallization of the bulk. To illustrate this Fig.~\ref{Fig4} displays the edge (upper left panel) and the bulk (upper right panel) site-resolved LDOS for $U=0.6<U_{TPT}$ and various temperatures across the CDW transition. Bottom panel shows the corresponding behavior of order parameter across the ribbon. Since the topology can only be canceled by strong enough CDW order, when the temperature grows the annealed disorder hinders the CDW, making the non-trivial topology more stable, i.e. increasing the size of the negative gap. If starting from $T=0$ the bulk had an {\it inverted} gap, no TPT upon heating is possible, cf. the upper right panel of Fig.~\ref{Fig4}. On top of that, reduction in the CDW order parameter leads to a more symmetric spectrum with respect to the chemical potential. However, at such small $U$ scattering does not introduce any low energy excitations due to gaped spectrum. It only contributes at higher energies by smearing out its peaked structure. The situation at the edges is different, where the edge states start to acquire finite lifetime due to scattering off the disorder. As a result they become flattened and more symmetric around the Fermi level, evolving towards an ``M-shape". Despite that they remain directional, one edge state connects the valence band with the conduction band from $K\rightarrow K^\prime$ and the other from $K^\prime \rightarrow K$. 

Free of the integer value constraint the CDW order parameter $d(y)$ starts to vary across the ribbon, cf. the bottom panel of Fig.~\ref{Fig4}. As the order-disorder transition is approached the $d(y)$ shows decays faster in the bulk then at the edges. It all stems from the presence of the edge states and their influence is twofold. Firstly, they lead to formation of an electron-rich and a hole-rich region at either end of the ribbon. In response, the localized particles redistribute from the former to the latter region in order to minimize the potential energy. It is important to note that despite $d(y)$ being symmetric with respect to the middle of the ribbon the total density monotonically decline across its width. The second effect of the edge states comes from their localization at a single sublattice. The additional/missing band at only one sublattice allows for stronger CDW order which enhances the redistribution of localized particles. As the edge state decays inside the ribbon so does the tail of the $d(y)$, highlighting the fact the two are connected.

\begin{figure}[ht]
\begin{center}
\includegraphics[width=0.5\textwidth]{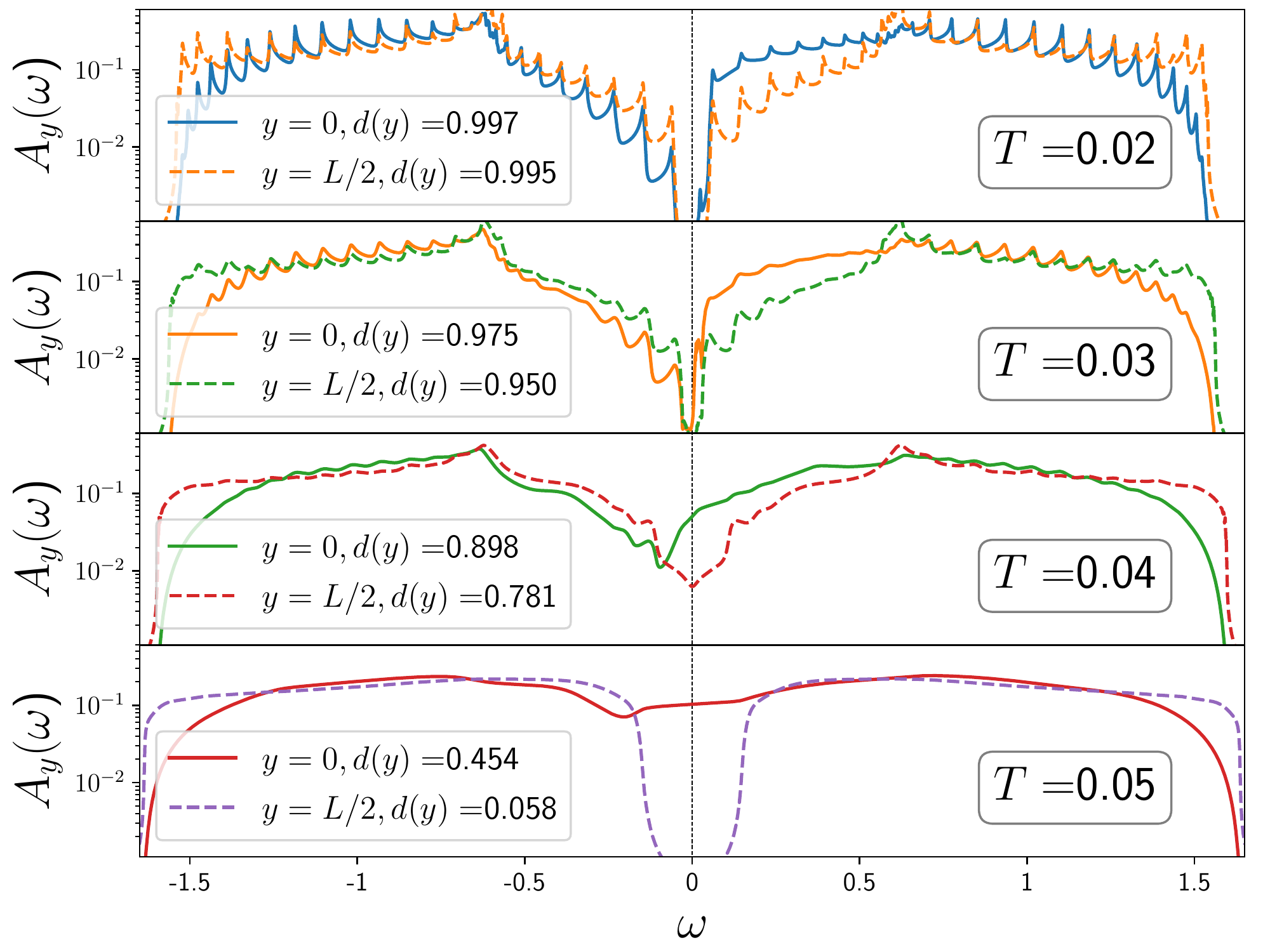}
\end{center}
\caption{Site-resolved LDOS for bulk (dashed line) and the edge (continuous line) for various temperatures at $U=1.2$. The legends on the left of each row display also the corresponding value of the order parameter. }
\label{Fig5}
\end{figure}

For interaction strengths $U>U_{TPT}$ the result of the competition between the topology and interactions becomes less straightforward. At $T=0$ the system starts with a trivial gap and upon heating it has to eventually end up in a disordered phase. This allows for a temperature induced TPT, when the CDW order falls below the threshold of stability of a trivial insulator. An example evolution upon heating of the edge (continuous line) and bulk (dashed line) total unit cell LDOS ($A_y(\omega)=0.5(A_{A,y}(\omega)+A_{B,y}(\omega)$) for $U=2.4$ is presented in Fig.~\ref{Fig5}. Because of the edge states, the $A_y(\omega)$ is not the same at sites $y$ equally distanced from the middle of the ribbon, but they are p-h reflected. The log-scale in Fig.~\ref{Fig5} is used to highlight the low energy part of the spectrum. 
The system starts as an ordered trivial insulator (OI) with well resolved CDW gap in the bulk and at the edges. As the temperature increases the order melts and trivial gap is reduced, cf. two top plots in Fig.~\ref{Fig5}. 
The comparison between $y=0$ and $y=L/2$ spectra reveals the (trivial) edge state contribution in form of a peak just below the upper edge of the bulk gap. The $y=L$ spectra (not shown) has similar peak located above the lower gap edge. 
At $T=0.03$ these peaks are located almost at the Fermi level, but the gap remains across all unit cells of the ribbon. 
This indicates that the system is approaching TPT from the trivial side.
Further increase of temperature melts the order below the threshold and the trivial gap is closed, as shown in the third panel from the top in Fig.~\ref{Fig5}.
At $T=0.04$ system becomes a TI phase as indicated by the edge turning metallic due to formation an asymmetric, in $\omega$, broad peak that bridges the previous CDW gap. This is a clear sign of presence of a topological edge state, cf. top right left in Fig.~\ref{Fig4}. Simultaneously, the low energy part of the bulk LDOS also undergoes an abrupt change. The sharp features, that sandwiched the CDW gap at lower $T$'s, are suddenly pushed away from Fermi level. The gap becomes softened due to tunneling of the edge state inside the ribbon, hence the low spectral weight subgap states. The fact that the tunneling of edge states is responsible for that was confirmed with a finite-size scaling analysis of metallicity of the bulk. It showed a steady decline of $A_{L/2}(0)$ with the system size and becoming fully suppressed around $L=30$. As a result of an edge state forming a large difference in the CDW order parameter between the edge and the bulk is observed, see inset of the $T=0.04$ panel in Fig.~\ref{Fig5}. This stems from the fact that the edge state is asymmetric around the Fermi level promoting additional increase in CDW order.

A topological state inducing a gapless bulk on the topologically non-trivial side of the TPT is a generic situation for finite size systems at $T\neq 0$, not discussed in previous studies of this model~\cite{PhysRevB.92.125102}. The mere presence of edge states in a topological non-trivial states softens the bulk gap by introducing a subgap bands, which tunnel into the bulk in a particular way. Despite most of the spectral weight being at the center of the edge state what tunnels inside is the part closes to the band edge.
When the {\it inverted} gap is narrow, as depicted for $T=0.04$, these subgap states are located very close to the Fermi level. The additional broadening of these states at finite temperatures, due to scattering, leads to eventual closing of the gap. From the third panel in Fig.~\ref{Fig5} one can see that these states are an order of magnitude smaller than any other features in spectra. It has to be stressed that the gapless bulk in this parameter range in a TI-CDW state is different from the ``standard`` gapless CDW (OC) in FKM~\cite{PhysRevB.76.205109}. They both originate from merger of subbands, but in the OC case these subbands are created solely by the disorder. Hence, the OC is stable also in the thermodynamic limit unlike this gapless TI-CDW. The part of the phase diagram it covers depends on the system size and as $L$ grows it collapse into a line. As the temperature and $U$ increase, the physics of the bulk start to dominate the tunneling effects and the gapless TI-CDW smoothly evolves into standard OC. 

In principle, on the topologically trivial side of the TPT a similar situation could take place as the low-energy peaks at the edges also tunnel inside the ribbon. Due to their small size, in contrast to their topological analogs, they decay much faster. Comparing the $T=0.02$ and $T=0.03$ in Fig.~\ref{Fig5} one can see the edge state peak being an order of magnitude smaller on the topologically trivial side then on the topologically non-trivial side of the transition. This makes a gapless bulk on topologically trivial side of the transition stable only at extremely small system size, at which point the parity of $L$ becomes of importance~\cite{PhysRevB.90.035116}.


Upon heating from $T=0.04$ to $T=0.05$ the CDW order further decreases and the {\it inverted} bulk gap grows. At $T=0.05$ it becomes well defined, but the remnants of the CDW are still present in all LDOSs across the ribbon in form of a small asymmetry around the chemical potential. This confirms the findings from Ref.~\cite{PhysRevB.92.125102} obtained using Hartree approximation, that even at $U>U_{TPT}$ a CDW survives inside a TI up to a certain temperature. The agreement between the Hartree approximation and full DMFT calculation proves the dominant role of the topology over interactions in this region of the $U-T$ parameter space. 
Further increase of temperature leads to complete suppression of order. The local LDOSs becomes that of a disordered TI, with symmetric LDOS and ``M-shaped" edge state band at low energies, cf. spectra for $U=0.6,T=0.02$ in Fig.~\ref{Fig4}. Once the system becomes maximally disordered further heating does not introduce any new behaviors, since temperature enters the mobile electrons LDOS only through the distribution of localized particles.

\begin{figure}[ht]
\begin{center}
\includegraphics[width=0.5\textwidth]{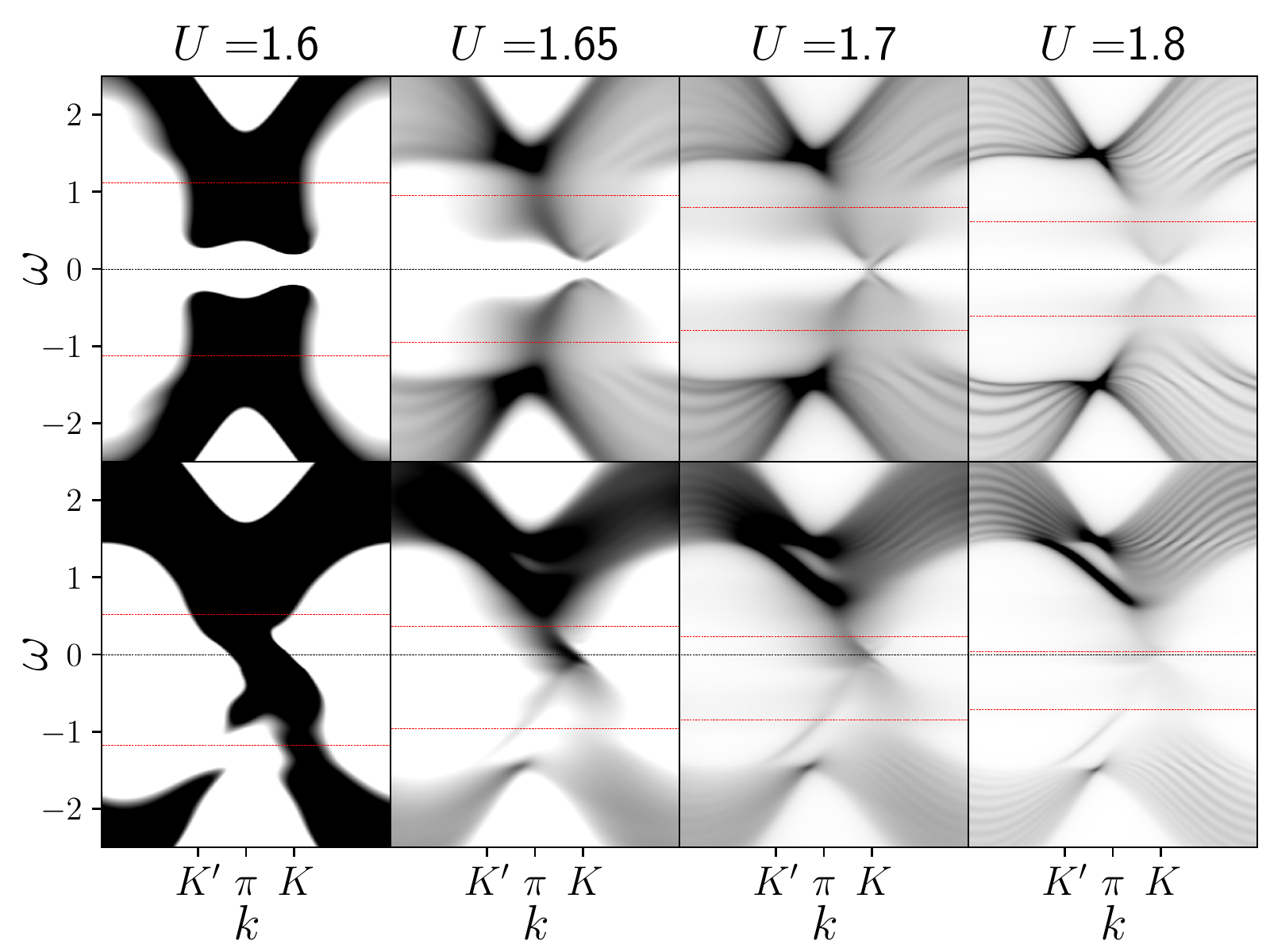}
\end{center}
\caption{Comparison of the dispersion along the unit cells at the edge (bottom row) and the middle (upper row) of the ribbon for various interaction strengths $U$ and $T=0.08$. The red dashed lines indicates the positions of maxima of the scattering rates around the Fermi level (black dashed line). }
\label{Fig6}
\end{figure}

The lack of the OC state at $U\approx U_{TPT}$ begs the question if it can be found at any point of the phase diagram and if so, how does the finite size of the system effect its properties. Away from the TPT one expects the system to follow the generic behavior of the FKM. Previous studies showed, that at any constant $T$ and upon increasing $U$ the system will eventually become a OC due to the cusp in its stability region, cf. phase diagram~\ref{phase_diag}, which reaches down to $T=0$ within the CDW dome~\cite{PhysRevB.76.205109}. The analysis of spectra at higher $U$  has the added benefit of having the scattering subbands separated from the clean system bands, making the study of the impact of finite size on various ordered phases easier. The Fig.~\ref{Fig6} shows a $U$-evolution of the dispersion along the edge (bottom row) and the dispersion along the middle of the ribbon (top row) at $T=0.16$. 

\begin{figure}[ht]
\begin{center}
\includegraphics[width=0.5\textwidth]{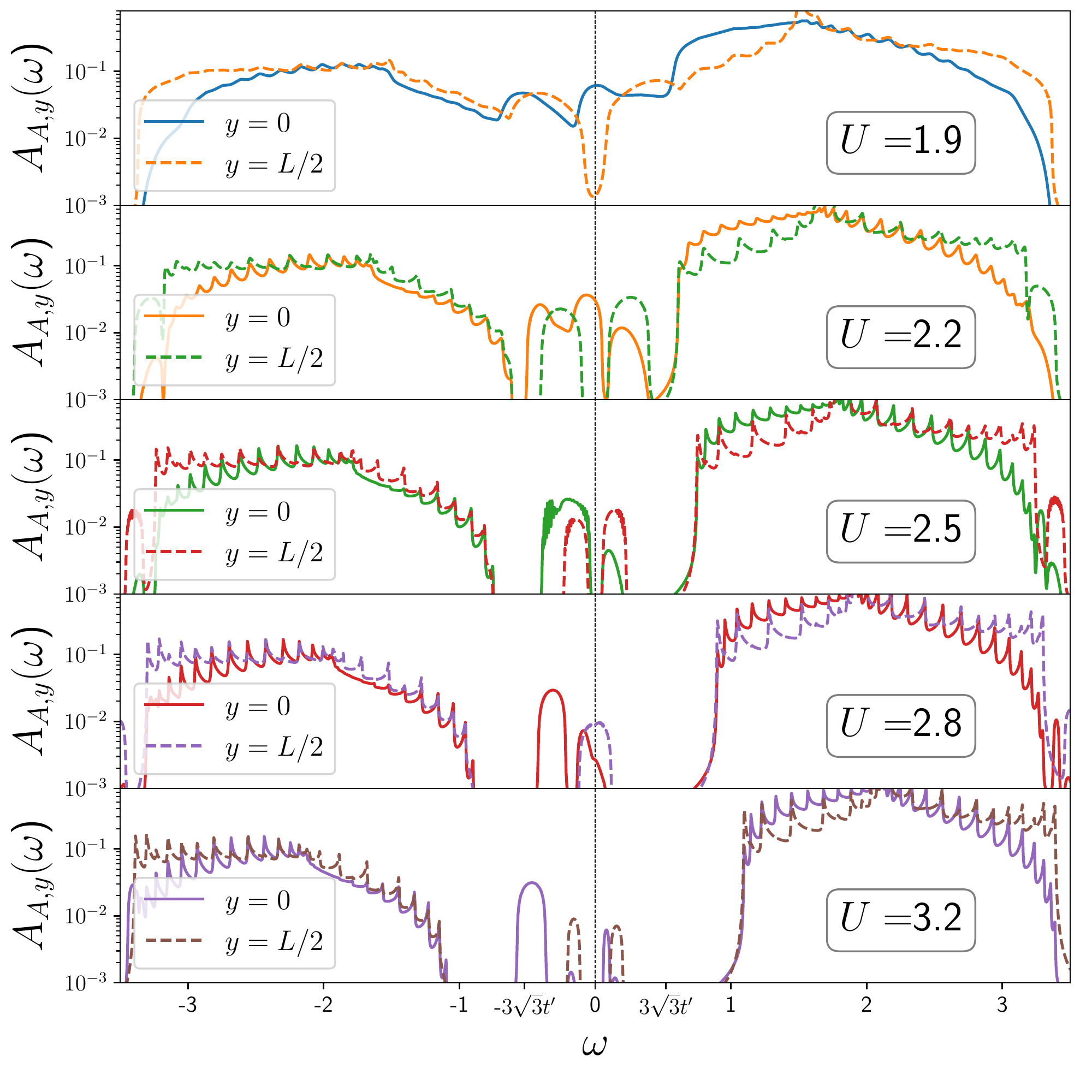}
\end{center}
\caption{Site-resolved LDOS for bulk (dashed line) and the edge (continuous line) for various interaction strengths $U$ at $T=0.08$. The logarithmic scale is used to highlight the interplay between the low energy scattering subbands.}
\label{Fig7}
\end{figure}

Starting at $U=1.6$ (leftmost column) the dispersion shows a (disoredered) TI, with a gapped bulk and broadened edge state crossing the Fermi level. The red dashed lines in Fig.~\ref{Fig6} indicate the location of local maxima of scattering rate. These maxima corresponds to the location of scattering subbands. In the bulk, they are located symmetrically above and below the Fermi level, reflecting the symmetry of the bulk unit cell. As the interaction is increased these peaks approach each other. At the same time the CDW order parameter increases. Equivalently, the annealed disorder is reduced. This can be inferred from the decrease in the broadening of the bulk bands, which start to recover a well resolved internal structucture for $U>1.7$. By extrapolating the position of the scattering subbands upon change of $U$, one can see that they will eventually cross (at $T=0.08$ it happens for $U=2.8$ , shown in Fig.~\ref{Fig7}) turing the bulk into a gapless CDW metal before reopening of the gap~\cite{PhysRevB.89.075104}.
For the parameters used in Fig.~\ref{Fig6} the closing of the {\it inverted} gap (at $U=1.7$) is not affecting the merging of the subbands, since the former takes place at much smaller $U$.

The situation at the edges is different, due to the fact they are locally doped away from half-filling. As shown in the bottom row the midpoint between the centers of the scattering subbands is below the Fermi level, with the subband on the hole-side ($\omega>0$) of the dispersion being closer to the Fermi level. As the system is being driven across the TPT this subband is pushed down to the Fermi level.
As a result the metallic nature of the edges remains despite the bulk becoming a topologically  trivial insulator. This ordered edge conductor phase (OEC) owes its metallic edges to the combined effect of scattering off annealed disorder and the existence of an additional bands at the zigzag edges of Haldane ribbon, which makes them locally doped.

At $U\approx 1.8$ the OEC phase does not show a clear gap in the bulk, neither does the system enters the gapless CDW phase (OC) from Ref.~\cite{PhysRevB.89.075104}. Instead, in this parameter range the metallicity of the bulk is just a finite size effect caused by a narrow bulk gap hybridizing with the metallic edges. This is similar to the situation discussed at the order-disorder transition at $U\approx U_{TPT}$, but this time the zero energy edge state is not of topological origin. As show in the top plot of Fig.~\ref{Fig7} the scattering subbands in the bulk LDOS (dashed line) are well separated and the residual spectral weight comes only from hybridization with the edge (continuous line). The low energy part of the spectrum at the edge does not show a broad asymmetric peak like for example at $U=0.6$ (see top left panel of Fig.~\ref{Fig4}), characteristic for an edge state. It is symmetric and narrow, typical for scattering subbands.
In order to better disentangle bulk from edge one has to look at larger $U$. The second panel in Fig.~\ref{Fig7} shows the same site-resolved local spectra for bulk and edge for $U=2.2$. At this interaction strength the CDW order opens a large enough gap to fully uncover the interplay between the subbands. The low energy part of the edge spectrum ($A_{A,0}(\omega)$) consists of two pairs of subband. One pair mirrors the subband structure of the bulk but with a reduced weight, clearly seen for the $\omega>0$ subband. Meaning they stem from hybridization. The other pair does not have an analog in the bulk spectra and the midpoint between the subbands is shifted to negative energies. This indicates they originated at the edge. The gap between them is filled by the subband that has tunneled from the bulk. The metallic nature of the edge remains due to the shifted position of the upper subband, whose upper edge is above the Fermi level. Following the nomenclature from Ref.~\cite{ActaPhysPolA.130.2} the bulk and edge is in some version of OI-X phases, the edge one being shifted in $\omega$, which is inferred from the subbands asymmetry.
Both the larger scattering subband and the larger band are on the same side of the Fermi level.

From this point on, increasing the interaction strength causes the system to alternate between metallic at $U=2.2,2.8$ and insulating state at $U=2.5,3.2$, due to exchange of the position of the bulk and the edge subbands. Firstly, at $U=2.5$ the edge subbands merge below the Fermi level which results in opening of the gap at the edges without closing it in the bulk. Later, at $U=2.8$ the bulk subbands merge turning the bulk, and due to hybridization the edge as well, metallic. Closer inspection of central peak in the $U=2.8$ edge spectra shows that the edge alone is already in the OI-Y phase but the upper sub-band is mixed with the contribution from the bulk. If this peak was purely a hybridization effect it would not have the same height as the bulk one, cf. the small negative energy part of the upper subband in the edge spectra. Finally, at $U=3.2$ the gap reopens. Both edge and bulk are in the OI-Y phase with larger scattering subbands on the opposite side of the Fermi level than the corresponding bands with higher spectral weight. Despite the subbands merging at different energies at the edges and inside the bulk, a phase with metallic bulk and gaped edges is not possible. Every time the bulk turns metallic it will tunnel into the edges. This overshadows any hypothetical MIT at the edges. For this reason metallic phase covers a larger part of the $U-T$ parameter space for the edges compared to bulk, cf. Fig.~\ref{phase_diag}.

\begin{figure}[ht]
\begin{center}
\includegraphics[width=0.5\textwidth]{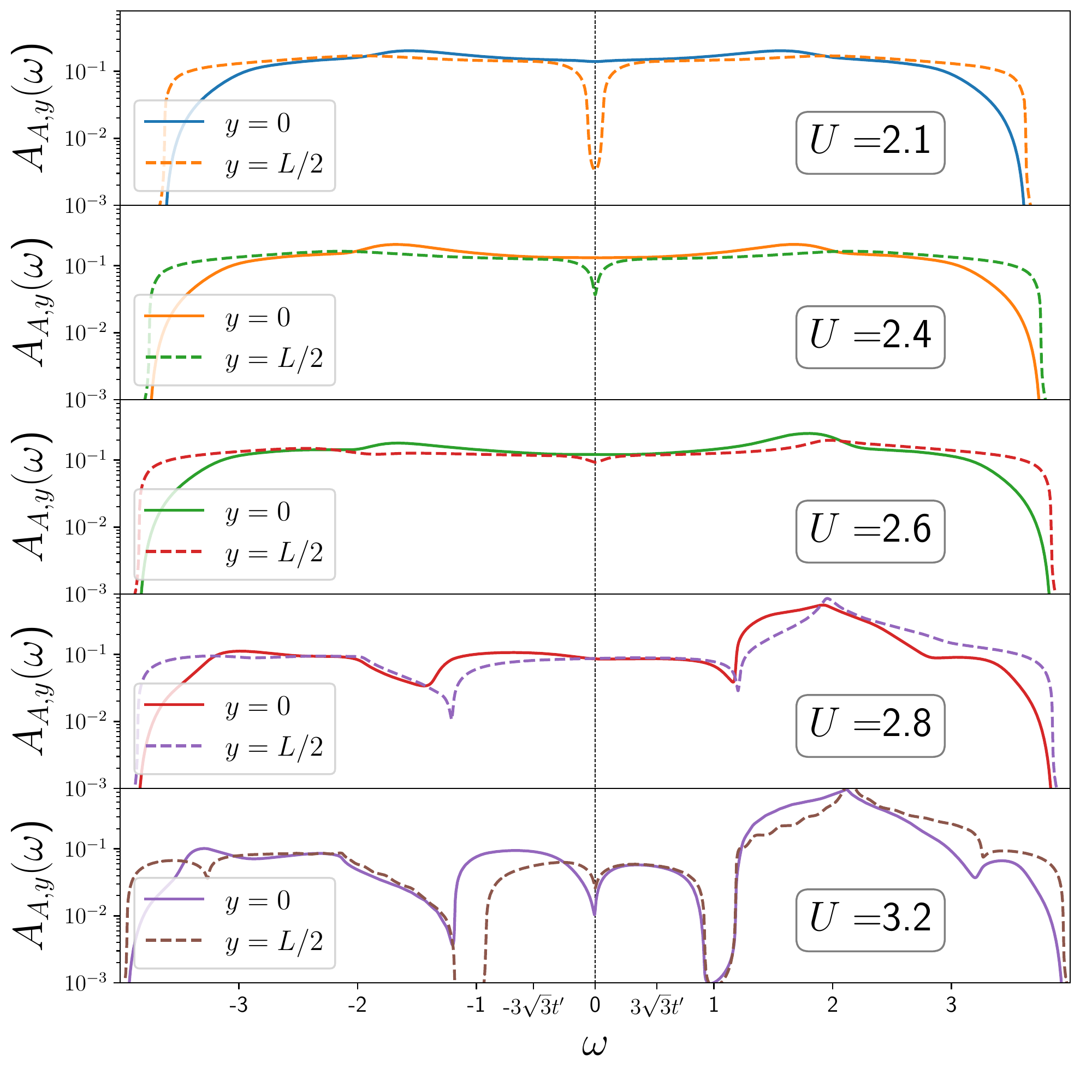}
\end{center}
\caption{The high temperature ($T=0.15$) evolution of site-resolved LDOS for bulk (dashed line) and the edge (continuous line) for various interaction strengths $U$. }
\label{Fig8}
\end{figure}

Following a similar evolution of LDOS with $U$ at higher temperatures one starts to observe a disordered metallic phase inside the bulk before order starts to form. Figure~\ref{Fig8} illustrates this for $T=0.15$ as the interaction changes from $U=2.1$ (top panel) to $U=2.4$ (second from the top).
The high temperature does not allow anymore for recognizing the topological state by comparing the bulk and edge spectra. Instead, to determine the topological state one has to anaylze the inverse of the zero frequency single-particle Greens function~\cite{PhysRevLett.105.256803,PhysRevB.92.125102}. In this cases it states that despite the metallic bulk the system is a topologically non-trivial state. The appearance of this phase is in agreement with previous Monte Carlo simulations~\cite{PhysRevLett.122.126601}, which showed such a topological gapless state (GTI) occupying a similar part of the phase diagram. The transition line to this phase is $T$-independent, because it appears in the part of the diagram where localized electrons do not form any order. The raise of the gapless bulk is purely due to formation of a Mott gap inside a topological gap. During this process the gap is smoothly reverted and the scattering off the maximal disorder introduce additional smearing of the gap edge. 
Hence the GTI region on the phase diagram has a maximal density of states at the Fermi level located in the middle.

\begin{figure}[ht]
\begin{center}
\includegraphics[width=0.5\textwidth]{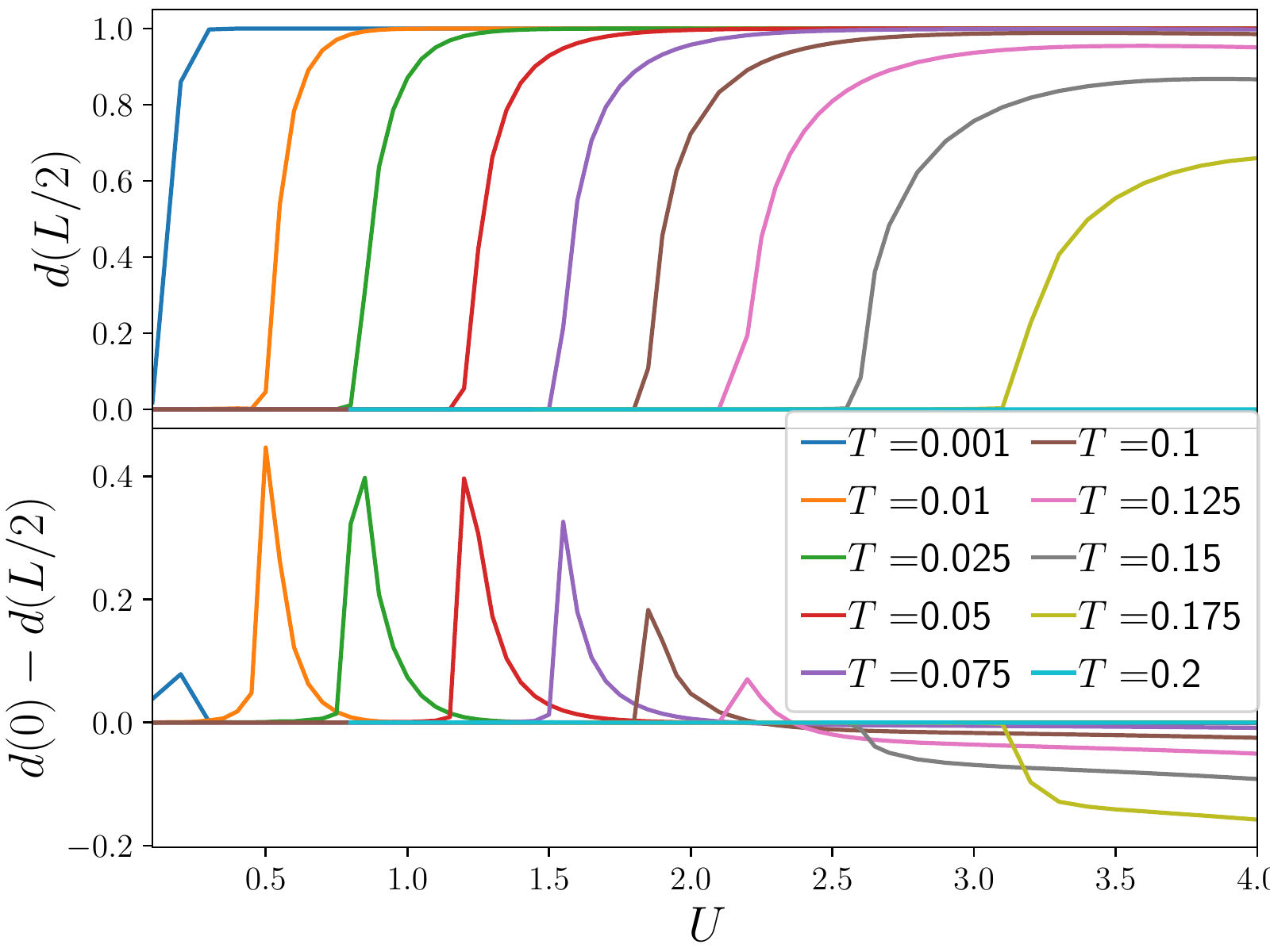}
\end{center}
\caption{Top panel: The bulk CDW order parameter $d(L/2)$ as a function of interaction strength $U$ for various temperatures. Bottom panel: the difference in the value of the CDW order parameter between the edge and the bulk as a function of $U$. In both panels the same colors correspond to the same temperatures.}
\label{Fig9}
\end{figure}

Further increase of $U$ triggers the formation of a CDW order. Initially, the bulk LDOS show small change in its shape from the disordered case. Eventually, it evolves towards a gapless CDW discussed above, as shown in forth panel from the top in Fig.~\ref{Fig8}. At this stage the edge and bulk display opposite asymmetries with respect to the Fermi level in their subbands. This reflects the fact that the two parts of the system are on different sides of their respective subbands merging point. In this region of $U-T$ space bulk shows stronger tendency to order then the edges, cf. the bottom panel of Fig.~\ref{Fig9}.  At smaller temperatures entering an ordered phase was more pronounced at the edges due to dominant role of the edge states. At higher temperatures the annealed disorder reduces this role up to the point where the physics of the OC phase becomes dominant. There the swap of positions of the scattering subbands at the edge introduce additional compensation of the unit cell polarization and the reduction in $d(0)$ follows. Making the CDW larger in the bulk then at the edge.
Similar change of the sign in $d(0)-d(L/2)$ is already present at lower temperatures when the system enters a state with opposite asymmetry of subbands in the bulk and at the, i.e. at $T=0.25$ and $U>4.5$. Change in sign of the difference of the CDW order parameter is not accompanied by any abrupt change in the the bulk CDW order parameter, top panel of Fig.~\ref{Fig9}, which has a monotonic behavior with $U$.

Lastly, taking a cut along even higher $T$ shows only an evolution from non-trivial to trivial disordered gaped bulk through a gapless bulk phase, which can be inferred from the edges becoming gaped at larger $U$.
Due to high temperature the system is not able to form an ordered state. As a result the temperature does not influence the spectra of mobile electrons and the transitions between different phases are solely due to the interplay between topology and interaction. Hence the transition lines in this part of the phase diagram are vertical, cf. Fig.~\ref{phase_diag}. This is consistent with the results from previous Monte Carlo calculations~\cite{PhysRevLett.122.126601} and a DMFT for homogenous phase on infinite lattice~\cite{PhysRevB.88.165132}. What differentiate the two approaches is the interpretation of the topological nature of the crossover regime.
Despite the lack of a CDW order the topological state becomes unstable at certain $U$ value at which point the system becomes a trivial (normal) homogeneous insulator. In contrast to the Monte Carlo results, before this transition the iDMFT show now change in topology of the system.

\section{Conclusions}

In this paper an analysis of the phase diagram of a Haldane ribbon with Falicov-Kimball interaction was presented. The results were obtained using iDMFT with fixed total densities of particles but with their distributions recalculated within each loop. This model allowed to analyze the influence including boundaries of a system in studying the interplay between topology and the CDW order formed due to interactions. Most of the results presented here were obtained for a ribbon with intermediate width, to allow for interference between edge states, but also to allow the middle part of the system to follow closely the thermodynamic limit behavior. Beside the presence of previously reported phases of the FKM on Haldane lattice the zigzag edges give rise to a topologically trivial ordered edge conductor. A phase whose conducting edges stem from locally doped boundaries and formation of scattering subbands. The locally doped nature of the zigzag edges was also shown to enhance the CDW order and induce small phase separation, by supplying an state for the mobile particles in one sublattice and removing at the other. Another consequence of zigzag edge was formation of gaped/gapless CDW phase that had different subband symmetry in the edge and bulk site-resolved spectra. Finally, it was shown that in the ribbon geometry a gapless topologial CDW state can exist in the region of parameter space that decays with the ribbons width. Thus, providing a reference point for determining the edge effects in bulk properties.

\section*{Acknowledgments}
The author would like to acknowledge M. Fabrizio, R. Lema\'{n}ski, and C. Mejuto Zaera for helpful discussions. This work was funded by the European Research Council (ERC) under the European Union's Horizon 2020 research and innovation programme, Grant agreement No. 692670 ``FIRSTORM''. 

\bibliographystyle{apsrev4-2}
\bibliography{mybiblio}
\end{document}